\documentclass[prd,preprint,a4paper,superscriptaddress,nofootinbib,11pt]{revtex4}
\usepackage{amsmath,amsfonts,amssymb}
\usepackage{txfonts}
\usepackage[T1]{fontenc}
\renewcommand{\d}{\mathrm{d}}

\begin{document}


\begin{center}
{\bf  \Large Entanglement Entropy Renormalization for the NC scalar field coupled to classical BTZ geometry}
 
 \bigskip
\bigskip

Tajron Juri\'c$^{a,}$ {\footnote{e-mail: tjuric@irb.hr}}
\; and \;  Andjelo Samsarov$^{b,}$ {\footnote{e-mail: samsarov@unica.it }}\\  
$^a$
Rudjer Bo\v{s}kovi\'c Institute, Bijeni\v cka  c.54, HR-10002 Zagreb,
Croatia \\[3mm] 
 
$^b$
Dipartimento di Matematica e Informatica, Universit\`{a} di Cagliari, viale Merello 92, 09123 Cagliari, Italy  and INFN, Sezione di Cagliari

\end{center}
\setcounter{page}{1}
\bigskip


\begin{center}
{\bf   Abstract}
\end{center}

    In this work, we consider a noncommutative (NC) massless scalar field coupled to the classical nonrotational BTZ geometry.
    In a manner of the theories where the gravity emerges from the underlying scalar field theory,
    we study the effective action and the entropy derived from this noncommutative model. In particular, the entropy is calculated by making use of the two different approaches,
    the brick wall method and the heat kernel method designed for spaces with conical singularity.
    We show that the UV divergent structures of the entropy, obtained through these two different methods, agree with each other. It is also shown that the same renormalization condition that removes the infinities from the effective action  can also be used to renormalize the entanglement entropy for the same system. Besides, the interesting feature of the NC model considered here is that
    it allows an interpretation  in terms of an equivalent system comprising of a commutative massive scalar field, but  in a modified geometry; that of the rotational BTZ black hole, the result that
    hints at a duality between the commutative and noncommutative systems in the background of a BTZ black hole.



\newpage
\section{Introduction}

That the  entropy can be assigned  to a black hole with the magnitude proportional to the horizon area was indicated for the first time in
\cite{bek1,bek2}  and later on this idea has been given a strong credibility \cite{hawking}.
In trying to understand why the black hole has the entropy and why is it proportional to the area,
two different approaches have been proposed in the $80's.$ One is due
to 't Hooft and is referred to as the brick wall model
\cite{gthooft}. In this method one considers a thermal bath of
particles propagating just outside of the horizon and calculates their
entropy. In the other, seemingly unrelated approach, the entropy is
calculated via introducing an auxiliary, but important concept of
reduced density matrix, which is obtained by tracing over the degrees
of freedom of a quantum field that reside inside the horizon
\cite{Bombelli:1986rw}. This later approach was in fact a seed which
triggered the whole stream of development that over the subsequent
years and in conjunction with the series of other important
contributions in the field gave rise to what may now be called the conical space or conical singularity approach to calculating the entanglement entropy \cite{dowker, Srednicki:1993im, Susskind:1993ws, Callan:1994py, frolov, fursaev, solodukhinfursaev}.
  
   Entropies obtained by either of these two approaches appear to be
   divergent quantities, which  naturally  raises the problem of their renormalization.
In this respect, it was suggested in \cite{Susskind:1994sm}
that the leading divergence in the entropy can be removed by the
   standard renormalization of Newton's gravitational constant.  
 Subsequently, the removal of the subleading divergent terms in the entropy by renormalizing the higher curvature couplings in the gravitational action was demonstrated in 
\cite{Solodukhin:1994yz, Fursaev:1994ea, Demers:1995dq}.

The conical singularity method is based on the simple replica trick first introduced in \cite{Susskind:1993ws}. It has been proven particularly elegant and powerful  in a number of situations of great physical interest.  
This method is particularly interesting when applied to black holes in trying to understand the dynamical origin of the entropy \cite{Callan:1994py, Solodukhin:1994yz, Solodukhin:1994st, Larsen:1995ax}.  
In that situation the entangling surface $\Sigma$ is the
Killing type of black hole horizon.
The characteristic feature that regular metrics with Killing type of
horizon have is that their Euclidean time direction is compact with
periodicity $ 2\pi \beta_{H}$, with $\beta_H$ being the inverse
Hawking temperature. Such property is dictated by the regularity
condition. The Euclidean time $\tau$ then plays the role of the angular
coordinate on the two dimensional disc which is perpendicularly oriented with
respect to the entangling surface $\Sigma$. In the small neighbourhood
of the surface $\Sigma$ the complete spacetime $E$ can be represented as
the direct product $E= \Sigma \times C$ of the entangling surface and
the two dimensional disk $C$. With $\tau$ and $\rho$ parametrizing the
disk $C$ and coordinates $z^i, i=1,...,d-2,$ parametrizing the surface
$\Sigma$, the spacetime $E$ is then assumed to be described by the static
and Euclidean metric of the general type
\begin{equation} \label{int1}
 ds^2 = g_{\mu \nu} dx^{\mu} dx^{\nu} = f(\rho) d\tau^2 + d\rho^2 +
 \gamma_{ij}(\rho, z^i) dz^{i} dz^{j}.
\end{equation}
Accordingly, the surface $\Sigma$ is determined by the condition $\rho
=0$ so that the expansion around the black hole horizon corresponds
to expanding around $\rho=0$.

One way to introduce a singularity into this otherwise regular spacetime
manifold is to displace the black hole out of its thermal equilibrium \cite{Hartle:1976tp, Susskind:1993ws, Callan:1994py}
by allowing it to have a temperature $T$ different from the
Hawking temperature $T_H$ $(T_H= 1/2\pi \beta_H)$. Naturally, such
displacement causes the appearance of a conical singularity which is
attached to the surface $\Sigma$ at the origin and has the deficit angle
$\delta = 2\pi (\beta_T - \beta_H)/\beta_T $. The regularity of the metric
can then be
restored by relaxing the black hole back to its thermal equilibrium at
the temperature $T=T_H$, resulting in the vanishing deficit angle, $\delta =0$.

Concerning the thermal entropy $S$ for a given field theoretical system at
some generic temperature $T$, one may recall that it is given in terms
of the partition function $Z_{\beta_T}$ of the field theoretical system in question,
\begin{equation} \label{int2}
 S= - \bigg( \beta_T \frac{\partial}{\partial \beta_T} -1 \bigg) \ln Z_{\beta_T}.
\end{equation}
This relation  in turn may be re-expressed in terms of the deficit
angle $\delta$ as
\begin{equation} \label{int3}
 S_{BH}= \bigg(2\pi  \frac{d}{d \delta} +1 \bigg) \ln Z_{\delta} \bigg |_{\delta =0}
\end{equation}
to give the Bekenstein-Hawking entropy as the thermal entropy evaluated at
the temperature $T=T_H$.


  On the other side, one may recall the concept of the reduced density
  matrix which appears to be very useful in the current
  context. Thus, for the system described by the pure quantum state
  ~ $|\psi \rangle$, ~ the density matrix is given by ~ $\rho_0 =  |\psi
  \rangle \langle \psi |$. If the system under consideration is further
  divided by some entangling surface $\Sigma$ in such a way  that a
  portion of the degrees of
  freedom is located within the surface, while the rest of them is located
  outside the surface, then a reduced density matrix  can be defined by
  tracing   ~ $\rho_0 $ ~ over the degrees of
  freedom  residing within the entangling surface.
 Even more, the reduced density 
matrix pertaining to
  any of the two artificially created subsystems can be defined and
  this can be done by
  tracing the density matrix ~ $\rho_0 $ ~ over the degrees of
  freedom that are located within the remaining of the two subsystems.
 Hence, if a given system is divided into two subsystems $A$ and $B$
  by some surface $\Sigma$, then the reduced density matrix for the
  subsystem $B$ is defined by tracing over the modes of the quantum
  field that reside in the subsystem $A,$
\begin{equation} \label{int4}
 \rho_B = \mbox{Tr}_A \; \rho_0
\end{equation}
and the entanglement entropy for the subsystem $B$ is given by  the von
Neumann's entropy as
\begin{equation} \label{int5}
 S_B = -\mbox{Tr} \; \rho_B \ln \rho_B.
\end{equation}

It is  known that the reduced density
  matrix ~$\rho$, properly normalized as $\hat{\rho} \equiv
  \frac{\rho}{\mbox{Tr} \; \rho},$  complies to the so called ``replica trick''
\begin{equation} \label{int6}
 S = -\mbox{Tr} \; \hat{\rho} \ln \hat{\rho} = \bigg( -\frac{d}{dn} + 1 \bigg)
 \ln \mbox{Tr} \; {\rho}^n  \bigg |_{n = 1},
\end{equation}
which provides the efficient way to calculate the entanglement entropy of
the quantum (field theory) system under consideration. Trace of the
$n$-th power of the reduced density matrix appears to be very
important quantity as the following line of arguments may plainly
show. 

 To begin with, one first has to note that calculating the expression $ \mbox{Tr} \; {\rho}^n $
corresponds to taking the path integral over field configurations that
are defined on the $n$-sheeted covering $E_n$ of the Euclidean
spacetime $E$ described by the metric (\ref{int1}). When calculating
this functional integral, one has to introduce a cut in the spacetime $E$ in order
to implement the 
boundary conditions obeyed by the quantum field. The cut is determined
by the points that lie on a half of the hyperplane $\tau = 0$.  Then,
in the process of taking the trace the value of the quantum field that lies on
one sheet immediately below the cut has to be identified with the
value of the quantum field that lies on the next sheet immediately above the cut (see e.g. \cite{Nishioka:2009un} for the nice graphical presentation).
In passing across the cut from one sheet to the other, fields are
glued analytically and two points which differ in angular variable by
$2\pi n$ are identified. Thus, we have a manifold on which two points
are identified
which differ in angular variable by some integral multiple of $2\pi$. Geometrically, this corresponds to a cone
with the deficit angle $ \delta = 2\pi (1-n)$.

   Re-expressing the replica formula (\ref{int6}) in terms of the
   deficit angle $\delta$, rather then in terms of $n$, gives
\begin{equation} \label{int7}
 S = -\mbox{Tr} \; \hat{\rho} \ln \hat{\rho} = \bigg( 2\pi \frac{d}{d\delta} + 1 \bigg)
 \ln  \mbox{Tr} \; {\rho}^n \bigg|_{\delta = 0}.
\end{equation}
If we are about to correlate the two entropies, the geometric entropy
(\ref{int7}) and the Bekenstein-Hawking entropy (\ref{int3}), then
by comparing the relations (\ref{int7}) and (\ref{int3}), a clear
physical interpretation of the quantity $ \mbox{Tr} \; {\rho}^n $  emerges,
namely it appears to be a partition function for the quantum field in some
gravitational background, $Z_n = \mbox{Tr} \; {\rho}^n $.

  Thereafter, the space $E_n$, which is a $n$-fold cover of the space
  $E$, is still described by the metric (\ref{int1}), with the only
  exception that the variable $\tau / \beta_H $ is no longer periodic
  variable with period $2\pi $, but instead  its period now is $2\pi n$.
  This space has a conical singularity attached to the surface
  $\Sigma$ at the origin, so that in the small neighbourhood of the
  entangling surface $\Sigma$ the space $E_n$ looks as a direct product
  $E_n = \Sigma \times C_n$ of the surface $\Sigma$ and a two
  dimensional cone with the deficit angle $\delta = 2\pi (1-n)$.

      Due to an abelian isometry which exists in the plane orthogonal to
      $\Sigma$ and which is  manifested by the periodicity with period
      $2\pi n$, it is possible to analytically continue above
      conclusions from integer $n$ to arbitrary noninteger  $\alpha$,
      so that  in the small vicinity of $\Sigma$
      it is possible to write $E_{\alpha} = \Sigma \times
      C_{\alpha}$. Correspondingly, the identification $Z_{\alpha} = \mbox{Tr} \; {\rho}^{\alpha} $
      may also be drawn.

At the same time,
the assumption about the spacetime as being described by a smooth manifold  at the energies of the order of the Planck scale 
 was being increasingly more challenged over the past few decades. Indeed, various different approaches to quantum gravity  in one or the other way point toward the necessity for 
  revising such description.
 One of
the approaches to quantum gravity takes this route and the spacetime is assumed to be noncommutative (NC)
at the microscopic level \cite{cones}. Such an assumption is not an arbitrary one, since general relativity and Heisenberg's
uncertainty principle together imply that the spacetime has a noncommutative structure \cite{dop1, dop2}. With this
prospect, different type of noncommutative spacetimes and their implications to physical models have
been analyzed in recent times \cite{RevNCnekrasov, RevNCszabo}.
Likewise, there have been
various attempts to construct noncommutative theories of gravity, noncommutative black hole solutions and
noncommutative quantum cosmology  \cite{wess1,wess2,ohl1,ohl2,mofat,ncg1,ncg2,ncg3,vor,Bastos:2007bg,Bastos:2009ae,LopezDominguez:2006wd}. In particular, it has been shown that the noncommutative
version of the BTZ black hole is described by a $\kappa$-deformed algebra \cite{brian1,brian2}. Similar $\kappa$-deformed algebras
have been found in the noncommutative description of Kerr black holes \cite{schupp} and certain noncommutative
versions of cosmology \cite{ohl2}. It thus appears that there is a certain element of universality in the appearance
of the $\kappa$-deformed algebras, as they occur in the noncommutative descriptions of various types of
classical geometries. Therefore it is of interest  to study the properties of black holes in the framework of
$\kappa$-deformed noncommutative systems.

In this paper we set up to
investigate the effects induced by  noncommutativity on the coupling of matter to gravity.
For that purpose, 
the dynamics of the scalar matter in the background of the BTZ
geometry \cite{banados,banados1}
has been sampled out as a working model for describing the matter coupled to gravity.
 On the other hand, as a conceptual framework for mimicking the noncommutative
nature of  spacetime at the Planck scale, the  $\kappa$-deformed Minkowski
spacetime has been envisioned as a convenient and sufficiently general
one, as explained above. 
Likewise, it is noteworthy that a non-smooth, grain-like nature of spacetime calls for a different types of symmetries which 
are compatible with it, since those embraced by the ordinary Poincar\'{e} are not.
Symmetries that underlie  $\kappa$-deformed systems are however embodied within 
the $\kappa$-deformed Poincar\'{e} algebra \cite{kappa1,kappa2,kappa3}, a
specific type of
quantum deformation of the Poincar\'{e} algebra.

The model we investigate is therefore based on the noncommutative  scalar field coupled to the classical BTZ
black hole background. 
It was first laid down in \cite{Gupta:2013ata} and in \cite{Gupta:2015uga} quasinormal modes are investigated in the light of searching for the effects  induced by noncommutativity.  
We bear on these results and expand the research along some novel
lines, with the strong emphasis on  two main aspects of the
problem. First one is the validity of the renormalization statement
{\footnote{Renormalization statement refers to the
    assertion that for renormalizing the entropy one does not need to
    invent a separate procedure, since the entropy renormalization can be carried out by the
    same redefinition of the couplings that served to renormalize the
    effective gravitational action. }} for the above described NC model in the spirit
of \cite{Susskind:1994sm} and the other is the evaluation of the
entropy for the same NC model. As for the latter case in
particular, the entropy for the NC model is calculated by using two
different methods, namely  the brick wall method \cite{gthooft}
 and the heat kernel method for conical spaces and the results, especially their UV divergent structures, are confronted with each other.
In trying to reach the stated milestones, we heavily rely on the heat kernel method designed for the spaces with conical singularity.
   The latter procedure fits neatly into the picture based on the long lasting idea due to Sakharov
    \cite{Sakharov:1967pk,Jacobson:1994iw,Visser:2002ew}, in which the gravity is not a fundamental property, but instead
    arises as a result of the quantum fluctuations that are due to the underlying quantum field theory. In the present case the underlying quantum field theory 
  is given by the NC scalar field in the classical BTZ background.



\section{Model for  NC scalar field in  BTZ background and 
      mapping to equivalent commutative model}

In \cite{Gupta:2013ata} a model was put forth that describes  the
coupling of the scalar particle to the metric of the form (in units
where $8G =1$)
\begin{equation}\label{btzmetric}
g_{\mu\nu}=\begin{pmatrix}
M - \frac{r^2}{l^2} &0&0\\
0&\frac{1}{\frac{r^2}{l^2}-M}&0\\
0&0&r^2\\
\end{pmatrix},
\end{equation}
and within the noncommutative  setting realized by the
presence of symmetries that are compatible with  $\kappa$-deformed
Poincar\'{e} algebra.
In this framework the noncommutativity enters through the scalar
field, which while describing matter, is treated as a noncommutative
object, compatible with the quantum (deformed) symmetry.
The gravity on the other hand is treated classically.
This approach therefore amounts to considering a noncommutative  scalar field
 coupled to the classical geometrical background produced by
 the spinless $(J=0)$ BTZ black hole with mass $M$. 

 

Salient features of the model are encoded within the radial equation of the form \cite{Gupta:2013ata,Gupta:2015uga}
\begin{equation}\label{eomradial}
r\left(M-\frac{r^2}{l^2}\right)\frac{\partial^2 R}{\partial r^2}+\left(M-\frac{3r^2}{l^2}\right)
\frac{\partial R}{\partial r}+\left(\frac{m^2}{r}-\omega^2\frac{r}{\frac{r^2}{l^2}-M}-a\beta\omega\frac{8r}{l^2}\frac{\frac{3r^2}{2l^2}-M}{\frac{r^2}{l^2}-M}\right)R=0,
\end{equation}
which is the radial component of the field equation for the NC scalar
field, and $a$ is the deformation parameter, $a=\frac{1}{\kappa}$,
(therefrom the phrase $\kappa$-deformation). It
 fixes the energy
scale at which NC effects are supposed  to start occurring. Most frequently it is
taken to be of the order of the Planck length.
  $l$ is related to the cosmological constant $\Lambda$ as $l =
  \sqrt{-\frac{1}{\Lambda}}$.
Furthermore, $\omega$ and $m$ are respectively the energy and the angular momentum
(magnetic quantum number) of the scalar particle.
The constant $\beta $ is the parameter determining the differential operator representation of the $\kappa$-Minkowski algebra. 
For more details see
\cite{Gupta:2013ata,Gupta:2015uga}.

 As a matter of fact, the field equation (Klein-Gordon equation) for the
 NC scalar field $\hat{\phi}$ can be rephrased in terms of the
 commutative reduction $\phi$ of $\hat{\phi}$, in which case the field equation takes the general form
\begin{equation} \label{1}
  ({\Box_g} + {\mathcal{O}}(a)) \phi = 0.
\end{equation}
By commutative reduction we mean the result of the limiting procedure $\hat{\phi} \longrightarrow
\phi,$ as $a \longrightarrow 0 $. Having said that,
Eq.(\ref{eomradial}) is in fact the radial component of the equation (\ref{1})
that governs the
dynamics of the commutative reduction $\phi$ of the NC scalar field
$\hat{\phi}$. As it can be seen, (\ref{1})  consists of two parts, the
first one being the standard Klein-Gordon operator for the geometry (\ref{btzmetric})
and the second one represents a novel contribution. It goes linearly with the NC scale $a$ and above all, introduces a new
physics. Note that this
$a$-dependent term in (\ref{1})  is  induced by the noncommutative nature of
spacetime at the Planck scale (or NC scale in general, whatever it be). It also gives rise to the corresponding term in the radial
equation (\ref{eomradial})  which scales
linearly with $a$. 

Using the substitution
\begin{equation} \label{2}
z=1-\frac{Ml^2}{r^2},
\end{equation}
Eq.(\ref{eomradial}) can be re-expressed  as
\begin{equation}
\label{eom}
z(1-z)\frac{\d^2 R}{\d z^2}+ (1-z)\frac{\d R}{\d z} + \left(\frac{A}{z}+B+\frac{C}{1-z} \right)R=0,
\end{equation}
where the constants $A,B$ and $C$ are
\begin{equation} \label{coefs}
A=\frac{\omega^2 l^2}{4M}+a\beta\omega, \quad B=-\frac{m^2}{4M}, \quad C=3a\beta\omega.
\end{equation}
 The equation  \eqref{eom}, together with the coefficients \eqref{coefs}, describes the dynamics of  massless NC scalar field with energy $\omega$ and angular momentum $m$, probing the geometry of a BTZ black hole with mass $M$ and vanishing angular momentum ($J=0$). Closer inspection of its form leads to an interesting observation. Namely, the analytical form of Eq.\eqref{eom}
   is exactly the same  as that of the equation of motion that governs the massive scalar field of mass $\mu'$, energy $\omega$ and angular momentum $m$, and probing the geometry of the rotational BTZ black hole with mass $M'$ and angular moment $J'$ (see Eq.(8) in ref.\cite{d1}).
   On purely technical grounds, the reason why this happened   is due to the fact that it was possible to absorb the  term with the noncommutative contribution right into those terms in the equation of motion that have already been present there in the absence of noncommutativity.
   As a result of this purely mathematical peculiarity, an interesting physical picture pops up. It appears that independently of the mass of the black hole whose geometry is being probed by the massless scalar particle,  this scalar particle will though 
   acquire the mass  in the presence of noncommutativity  and  will simultaneously undergo some type of back-reaction, deforming the geometry through which it propagates. In particular, the latter feature  pertains to  both, the modification of the black hole mass as well as the change in the very nature of the geometrical background
    that is being probed, forcing it to alter from a non-rotational  into
    a rotational one (with the angular momentum $J$ different from
   zero). In accordance with the above observations, a novel perspective may
   be given to spacetime noncommutativity. Not only that it may be
   given a role of the driving force that lies behind the mass generating mechanism, but  it may also be responsible or may give rise to certain back-reaction effects.

   
 Thereby, since we have an equivalence between  two equations of motion, which  pertain to  two completely different physical situations, a question naturally arises as to whether it is possible to find some kind of mathematical correspondence between them. The answer is positive. Namely, it appears that it is  possible  to
     find a direct mapping between the case considered here, that is
     NC massless scalar field in the non-rotational BTZ background,
     and  the physical setting where the ordinary massive scalar field
     probes a BTZ geometry with nonvanishing angular momentum. In what
     follows, the latter setting we shall refer to  as the fictitious
     one (see ref.\cite{d1} for the explicit expressions that correspond to
     this fictitious situation).

        Therefore, 
      by comparing the constants $A,B$ and $C$, appearing in \eqref{eom}, with the appropriate constants from the reference \cite{d1}, we get the following set of conditions
\begin{equation}\begin{split} \label{8}
&A=\frac{\omega^2 l^2}{4M}+a\beta\omega=\frac{l^4}{4(r'^2_{+}-r'^2_{-})^2}\left(\omega r'_{+}-\frac{m}{l}r'_{-}\right)^2=A'   \\
&B=-\frac{m^2}{4M}=-\frac{l^4}{4(r'^2_{+}-r'^2_{-})^2}\left(\omega
  r'_{-}-\frac{m}{l}r'_{+}\right)^2=B' \\
&C=3a\beta\omega=-\frac{\mu'}{4}=C',   
\end{split}\end{equation}
with $\; r'_+, r'_- \;$ being the outer, i.e. inner radius of the
equivalent spinning BTZ black hole, respectively.

Note that  the radii pertaining to the spinless BTZ are
respectively given by $r_+ = l\sqrt{M}$ and  $r_- =0 $. In addition, to
keep the notation as simple as possible, we omit the superscript from
$\mu',$ so that hereafter we have $\mu' \equiv \mu$. 

Furthermore, since \cite{banados}
\begin{equation}
M'=\frac{r'^{2}_{+}+r'^{2}_{-}}{l^2}, \quad  J'=\frac{2r'_{+}r'_{-}}{l},
\end{equation}
we can express the parameters of the commutative fictitious situation completely in terms of the parameters defining the NC case  we analyze here,
\begin{eqnarray} \label{3}
 M' & = &  f_1 (a,M), \\
 J' & = & f_2 (a,M), \\
  {\mu}' \equiv \mu & = & f_3(a,M).
\end{eqnarray}
This mapping is similar to the one obtained in \cite{veza}, where the
analogy between the NC version of the Schwarzschild black hole and the commutative Reisner-Nordstrom black hole was drawn. 

Turning back to the conditions (\ref{8}),
it is possible to solve them and to get the fictitious parameters $M'$
and $J'$  in a closed form. This looks as
\begin{equation} \label{6}
 \frac{1}{M'} \frac{1}{{\omega}^2 l^2 + m^2 - 2\omega m \frac{J'}{M'}}
 = {\bigg( \frac{1}{M} + \frac{4a\beta \omega}{{\omega}^2
 l^2 - m^2} \bigg)}^2 \frac{1}{\frac{{\omega}^2 l^2 + m^2}{M} +
 4a\beta \omega},
\end{equation}
where the ratio $J'/M'$ appearing in (\ref{6}) is given by
\begin{equation} \label{4}
 \frac{J'}{M'} = \frac{\sigma \lambda {\gamma}^2 - \sqrt{{\gamma}^2
 {\sigma}^2 - \frac{{\gamma}^2 {\lambda}^2}{l^2} +
 \frac{1}{l^2}}}{{\gamma}^2 {\sigma}^2 + \frac{1}{l^2}}.
\end{equation}
Remaining abbreviations $\gamma, \lambda$ and $\sigma,$ appearing in the last two expressions, are listed
as follows
\begin{eqnarray} \label{5}
 \gamma & \equiv & \frac{\frac{1}{M} + \frac{4a\beta \omega}{{\omega}^2
 l^2 - m^2}}{\frac{{\omega}^2 l^2 + m^2}{M} + 4a\beta \omega}, \\
 \lambda & \equiv & {\omega}^2 l^2 + m^2, \\
  \sigma & \equiv & 2 \omega m.
\end{eqnarray}
 The mass $M'$ and the angular momentum $J'$ of the equivalent black hole can be expressed explicitly within the first order in the deformation,
\begin{eqnarray} \label{mprimejprime}
  M' & =& M \bigg[ 1 + 4a\beta \omega M \bigg( \frac{1}{\lambda} -\frac{2}{{\omega}^2 l^2 - m^2}
    +\frac{l^2}{\lambda^{2}} \frac{2\sigma^{2} \lambda^{2} -\sigma^{2} l^{2} + \lambda^{2} }{\sigma^{2} l^{2} + \lambda^{2}} \bigg( \frac{1}{{\omega}^2 l^2 - m^2}
      - \frac{1}{\lambda} \bigg) \bigg) \bigg], \nonumber \\
  J' &=& 4 a\beta \omega M^{2} \frac{l^{2}}{\lambda \sigma}   
      \frac{2\sigma^{2} \lambda^{2} -\sigma^{2} l^{2} + \lambda^{2} }{\sigma^{2} l^{2} + \lambda^{2}} 
   \bigg( \frac{1}{{\omega}^2 l^2 - m^2}
         - \frac{1}{\lambda} \bigg).
\end{eqnarray}
In order to understand the physical meaning behind the above
equivalence, it should be noted that when the noncommutative parameter $a$ (NC scale) goes to zero,
the parameters of the two situations coincide with each other. In
particular, the ratio $J'/M'$ goes to $0,$ as expected. Moreover, it should be noted
that the
right hand sides of the relations (\ref{6}) and (\ref{4}), beside
depending on $a$ and
the 'old' parameter $M$, also depend on the quantum state of the
scalar field through their dependence on the quantum numbers $m$ and
$\omega.$ 
For example, the relation (\ref{mprimejprime}) implies that the scalar particle with zero orbital angular momentum $(m=0)$ cannot change the spin of the black hole, although it can change its mass. 
We may additionally portray the whole situation   by saying that the more energetic the
incoming scalar field is, the
more intensive is its impact on the geometry it is probing.  This
observation speaks in favour of the above posed assertion that the noncommutativity of the scalar field generates, possibly through some back-reaction, the additional mass and angular momentum of the system with the fictitious black hole.  

To summarize this part, the picture that has emerged so far is the following.
The dynamics of the NC massless scalar field in the geometry (\ref{btzmetric})
is described by the equation (\ref{1}), where $\Box_{g}$ is the
Klein-Gordon operator for the metric (\ref{btzmetric}).
 Likewise, the dynamics of the massive commutative scalar field in the
 geometry  
\begin{equation}\label{eqbtzmetric}
g_{\mu\nu}=\begin{pmatrix}
 M' - \frac{r^2}{l^2} &0& \frac{-J'}{2}\\
 0&\frac{1}{\frac{r^2}{l^2} + \frac{J'^2}{4 r^2} -M'}&0\\
\frac{-J'}{2} &0& r^2\\
\end{pmatrix},
\end{equation}
is described by the equation
\begin{equation} \label{7}
  ({\Box_{g'}} - {\mu'}^2) \phi = 0,
\end{equation}
where $\Box_{g'}$ is the
Klein-Gordon operator for the metric (\ref{eqbtzmetric}).
As demonstrated above, these two different physical situations are mathematically equivalent,
due to the fact that Eq.(\ref{1}) can be rewritten and reduced
to the form (\ref{7}).

As already indicated,
such direct mathematical correspondence enables one to give a physical
interpretation to  NC effects. Consequently, we may say that probing a BTZ ($M$,$J=0$) black hole
with a massless NC scalar field appears to be equivalent to a
fictitious commutative setting where this same scalar field (its
commutative reduction, to be more precise)  acquires the mass and
simultaneously modifies the  geometry through which it propagates, possibly
through some mechanism of back-reaction. We know that the gravitational
background influences the particle. However, it may  be that 
the opposite is also true, with the grain-like, noncommutative nature of spacetime
providing a sufficiently suitable medium/agent for making something like this come true.

In the next three sections the focus will be on the entropy issue for
 the NC model just described. In  section III the entropy will be
 calculated by using the brick wall method. After reviewing in section
 IV the essentials
 of the heat kernel method for the spaces with conical singularity, in
 section V the entanglement entropy for the same NC model will be discussed.
However, when carrying out the calculations within the latter framework, one has to work with the
 Euclidean metric. Therefore,
before commencing the analysis of sections IV and V, it is necessary to make an analytic
transformation of the Lorentzian metric into the Euclidean one. We do
this by changing the real variables of time $t$ and angular momentum
$J'$ into 
\begin{equation}
 \tau = it, \quad J_E = -iJ',
\end{equation}
leading to the metric
\begin{equation} \label{euclideanmetric}
  ds_E^2 = \bigg( \frac{r^2}{l^2} - \frac{J_E^2}{4r^2} - M'
  \bigg)d\tau^2 + \frac{dr^2}{\frac{r^2}{l^2} - \frac{J_E^2}{4r^2} - M'}
  + r^2 \bigg( d\varphi - \frac{J_E}{2r^2 } d\tau \bigg)^2.
\end{equation}
The idea here is to carry out the calculations in the Euclidean setting
and then after the final result is reached, one again switches back to
the Lorentzian form by using the same transformations.
One more thing that one has to keep in mind is that the conical singularity method is an off-shell method,
meaning that the metric for which the entanglement entropy is
calculated does not necessarily need to be a solution to any field equations.
Even if it is, one inserts the specific metric into the formulas only
after the final formula for the entropy is derived.



\section{Entropy and its divergent structure  for the NC scalar field in
  the classical BTZ background from the brick wall method }

The method for calculating the entropy of the black hole by using ``brick-wall model'' was introduced in the seminal paper  \cite{gthooft} by 't Hooft. For the case of the BTZ black hole the method has been applied in \cite{kim} and in \cite{Ho:1998du} it was used to study the rotational BTZ case.
 By following the same line of arguments as in these papers, we find from Eq.\eqref{eomradial} that  the $r$-dependent radial wave 
number has the following form \cite{Gupta:2013ata}
\begin{equation}\label{k}
k^2 (r,m,\omega)=-\frac{m^2}{r^2\left(\frac{r^2}{l^2}-M\right)}+\omega^2\frac{1}{\left(\frac{r^2}{l^2}-M\right)^2}+a\beta\omega\frac{8}{l^2}\frac{\frac{3r^2}{2l^2}-M}{\left(\frac{r^2}{l^2}-M\right)^2}.
\end{equation}
In obtaining the last expression, we have used the WKB approximation, which assumes the ansatz of the form $R(r)=\text{e}^{i\int k(r)\text{d}r}$. According to the semi-classical quantization rule, the radial wave number is quantized as
\begin{equation}
\pi n=\int^{L}_{r_{+}+h} k(r,m,\omega)\text{d}r
\end{equation}
where the quantum number $n>0$ and the angular momentum quantum number
$m$ should be fixed so that  $k^2 (r,m,\omega) > 0.$ Note that
$n\equiv n(m,\omega).$ Alongside, $h$ and $L$ are the ultraviolet
and infrared regulators, respectively.
In the subsequent calculation for the free energy and entropy we shall take the limit
$L\rightarrow\infty$ at the end of calculation and set $h\approx 0$. In ref.\cite{Gupta:2013ata}
the leading, i.e. the most dominant term was calculated. Here, we shall extend the calculation and isolate all UV divergent terms. Special focus will be on calculating the next to leading term in the entropy and free energy.

  The total 
number $\nu$ of single particle solutions with energy not exceeding $\omega$ is given by
\begin{equation} \label{nu}
\nu \equiv \nu (\omega) =\sum^{m_{0}}_{m= -m_{0}}n(m,\omega) =
 \int^{m_{0}}_{-m_{0}}\text{d}m ~n(m,\omega) = \frac{1}{\pi}\int^{m_{0}}_{-m_{0}}
\text{d}m\int^{L}_{r_{+}+h} k(r,m,\omega)\text{d}r,
\end{equation}
where
$
 m_0^2 = \frac{\omega^2 l^2}{z} + a\beta \omega \frac{8M}{z}
\frac{z+ 1/2}{1-z}
$
 is fixed by the requirement $k^2 (r,m,\omega) >
0. \;$ Note that for reaching the conclusion on the value of $m_0,$  the change of the variable (\ref{2}) has been made
in the above integral, so that $k(r) \text{d}r = \kappa(z) \text{d}z,$
 for an appropriate function $ \kappa(z)$ (see (\ref{F}) for its explicit form). Accordingly, the bounds of
 integration in (\ref{nu}) are  changed to $z_h = 1 -
 \frac{Ml^2}{{(r_+ + h)}^2} $ and $z_L = 1 -
 \frac{Ml^2}{L^2}. $

The free energy at the inverse temperature $\beta_{T}$ of the black hole is 
\begin{equation}
\text{e}^{-\beta_{T}F} = \sum_{\nu}\text{e}^{-\beta_{T}E}=\prod_{\nu}\frac{1}{1-\text{e}^{-\beta_{T}E}},
\end{equation}
which after taking the logarithm on both sides leads to
\begin{equation}\begin{split}
\beta_{T}F&=\sum_{\nu}\text{ln}\left(1-\text{e}^{-\beta_{T}E}\right)=\int\text{d}\nu\text{ln}\left(1-\text{e}^{-\beta_{T}E}\right)\\
&=-\int^{\infty}_{0}\text{d}E\frac{\beta_{T}\nu(E)}{\text{e}^{\beta_{T}E}-1},
\end{split}\end{equation}
where the last line is obtained by the partial integration.
For this, we find the free energy $F$ as
\begin{equation}
F=-\frac{1}{\pi}\int^{\infty}_{0}\frac{\text{d}\omega}{\text{e}^{\beta_{T}\omega}-1}\int^{z_L}_{z_h}\text{d}z\int^{m_{0}}_{-m_{0}}\text{d}m\ \ \kappa(z,m,\omega).
\end{equation}
or more explicitly,
\begin{equation} \label{F}
F=-\frac{1}{2\pi}\int^{\infty}_{0}\frac{\text{d}\omega}{\text{e}^{\beta_{T}\omega}-1}\int^{z_L}_{z_h}\text{d}z\int^{m_{0}}_{-m_{0}}\text{d}m\
\ \sqrt{\frac{1}{z(1-z)} \bigg[ -\frac{m^2}{M} + \frac{\omega^2
    l^2}{Mz} + 8a\beta \omega \frac{z+ \frac{1}{2}}{z(1-z)}   \bigg]}.
\end{equation}
The integration over $m$ can be performed exactly and it yields
\begin{equation} \label{F1}
F=-\frac{1}{4}\int^{\infty}_{0}\frac{\text{d}\omega}{\text{e}^{\beta_{T}\omega}-1}\int^{z_L}_{z_h}\text{d}z
 \sqrt{\frac{1}{z(1-z)}} \frac{\omega^2
    l^2}{\sqrt{M}z} 
   -\frac{1}{4}\int^{\infty}_{0}\frac{\text{d}\omega}{\text{e}^{\beta_{T}\omega}-1}\int^{z_L}_{z_h}\text{d}z
 \sqrt{\frac{1}{z(1-z)}}
    8a\beta \omega \sqrt{M} \frac{z+ \frac{1}{2}}{z(1-z)}. \nonumber
\end{equation}
After carrying out the integrations over $z,$ one gets
\begin{equation}
 F=-\frac{1}{4}\int^{\infty}_{0}\frac{\text{d}\omega}{\text{e}^{\beta_{T}\omega}-1}
    \frac{\omega^2 l^2}{\sqrt{M}} (-2)  \sqrt{\frac{1-z}{z}} \; \Bigg |_{z_h}^{z_L}  
   -\frac{1}{4}\int^{\infty}_{0}\frac{\text{d}\omega}{\text{e}^{\beta_{T}\omega}-1}
        8a\beta \omega \sqrt{M}  \frac{4z-1}{\sqrt{z(1-z)}} \; \Bigg |_{z_h}^{z_L} . 
\end{equation}
Next we extract all divergent contributions to the free energy. For that purpose both terms in (\ref{F1}) are expanded in  the brick wall cutoff $h$. Keeping all divergent terms in the first term  gives
\begin{equation}
  (-2)\sqrt{\frac{1-z}{z}} \; \Bigg |_{z_h}^{z_L}  
   =  \sqrt{\frac{2l \sqrt{M}}{h}} \bigg( 1- \frac{h}{l \sqrt{M}} + O(h^2) \bigg), \nonumber
\end{equation}
while the second term leads to 
\begin{equation}
  \frac{4z-1}{\sqrt{z(1-z)}} \; \Bigg |_{z_h}^{z_L}
   =  \bigg( 1- \frac{8h}{l \sqrt{M}} \bigg) \sqrt{\frac{l \sqrt{M}}{2h}} 
     {\bigg( 1- \frac{2h}{l \sqrt{M}} \bigg)}^{-1/2} 
      = \sqrt{\frac{l \sqrt{M}}{2h}} \bigg( 1+ \frac{h}{l \sqrt{M}} + O(h^2) \bigg). \nonumber
\end{equation}

This means that only the leading term is divergent, while all other terms, including the next to leading term, are UV finite. Therefore, the total divergent part of the free energy is given as
\begin{equation}
F=-\frac{l^{\frac{5}{2}}}{(M)^{\frac{1}{4}}}\frac{\zeta(3)}{\beta^3_{T}}\frac{1}{\sqrt{2h}}-2a\beta\frac{(M)^{\frac{3}{4}}\sqrt{l}}{\sqrt{2h}}\frac{\zeta(2)}{\beta^2_{T}},
\end{equation}
which is the exact result in the sense of the  WKB method and $\zeta$ is the  Riemann zeta function.

 The corresponding divergent structure of the entropy for the NC massless scalar field  is calculated using the relation 
$S=\beta^2_{T}\frac{\partial F}{\partial \beta_{T}}$ and it accordingly amounts to
\begin{equation}\begin{split} \label{entropy}
S&=3\frac{l^{\frac{5}{2}}}{(M)^{\frac{1}{4}}}\frac{\zeta(3)}{\beta^2_{H}}\frac{1}{\sqrt{2h}}+4a\beta\frac{(M)^{\frac{3}{4}}\sqrt{l}}{\sqrt{2h}}\frac{\zeta(2)}{\beta_{H}}\\
&=S_{0}\left(1+\frac{4}{3}a\beta\frac{M}{l^2}\frac{\zeta(2)}{\zeta(3)}\beta_{H}\right),
\end{split}\end{equation}
where $S_{0}$ is the undeformed entropy for BTZ at the Hawking temperature $\beta_{T}=\beta_{H} =\frac{2\pi l^2}{r_{+}}$. The result for $S_0$  coincides with the result of \cite{kim}, while the additional term that scales linearly with $a$ is a consequence of the presumed noncommutative nature of spacetime.

\subsection{Dimensional analysis}

We work in units $\hbar = c = k_B = 8G =1$.
From the well known relations $E=k_B T,$ $E=mc^2,$ $E= \hbar \omega,$ $p=\frac{\hbar}{\lambda},$ $l=ct,$ $\beta_T = \frac{1}{k_B T}$ we have the following dimensional relationships,
\begin{align} \label{diman}
[E] &=[T]=[m]=[\omega] = [p], \nonumber \\
[p] &= [l]^{-1} = [E],  \nonumber \\
[l] &= [t],   \\
[\beta_T] &= [T]^{-1} = [E]^{-1} = {([l]^{-1})}^{-1} =[l].   \nonumber
\end{align}
The more explicit relation  between the brick wall cutoff $h$ and the invariant proper length $\epsilon$ between the horizon and the brick wall  
is visible \cite{Ho:1998du} from $\epsilon = \int_{r_+}^{r_+ + h} \sqrt{g_{rr}} \; \text{d} r.$
Namely,
$$
 \epsilon = \int_{r_+}^{r_+ + h} \frac{\text{d} r }{\sqrt{\frac{r^2}{l^2} -M}}  
$$
leads to the relation 
\begin{equation}
  r_h = r_+ + \frac{\pi \epsilon^2}{\beta_H} = r_+ + h.
\end{equation}
From this and (\ref{diman}),
a dimensional analysis gives $[h] =\frac{[\epsilon^2]}{[\beta_H]} = \frac{[\epsilon]^2}{[l]}.$ Since
$h$ has the dimension of length, $[h]=[l],$ it follows $ [l] = \frac{[\epsilon]^2}{[l]}, $  that is, the geodesic invariant distance $\epsilon$ between the horizon and an imaginary brick wall
also has the dimension of length, $ [\epsilon] =[h]=[l]$.

  Expressed in terms of the geodesic invariant distance cutoff $\epsilon$
and upon utilising the area formula $A (\Sigma)= 2\pi r_{+},$ the entropy (\ref{entropy})  takes the form
\begin{equation} \label{entropymain}
  S=\frac{3}{8\pi^3} \zeta(3) \frac{A(\Sigma)}{\epsilon} \left(1+\frac{4}{3}a\beta\frac{M}{l^2}\frac{\zeta(2)}{\zeta(3)}\beta_{H}\right),
\end{equation}
manifesting the area law for the NC model.

\section{Heat kernel method for spaces with conical singularity}

As we have already noted, the black hole spacetime $E$ in the vicinity of the
horizon $\Sigma$, which  plays the role of the entangling surface here, may be
represented by the direct product of the compact surface $\Sigma$ and
a two-dimensional disc $C$, $E = \Sigma \times C$, with the Euclidean
time $\tau$ playing the role of the angular coordinate on the disc, with the period $2\pi \beta_H$. The
conical singularity may then be introduced into this spacetime by virtue of
displacing the black hole out of its thermal equilibrium, which is effectively
  achieved by allowing the horizon temperature $T$ to depart from the
Hawking temperature $T_H$ by some small amount, thus resulting in the Euclidean spacetime $E_{\alpha} $ with a conical singularity. The conical
singularity  introduced in such way  is then
located  at the horizon and the spacetime $E_{\alpha} $ in the neighbourhood
of  the singular horizon surface then looks as
$E_{\alpha} = \Sigma \times C_{\alpha}$, where $C_{\alpha} $ is the
two-dimensional cone with the angular deficit $\delta = 2\pi (1- \frac{T}{T_H}) \equiv 2\pi (1- \alpha )$.
This means that in the vicinity of the horizon the metric is still described by (\ref{int1}),
except for the fact that the Euclidean time $\tau$ is now a periodic variable with period
 $2\pi \beta_T$.

   Moreover, in theories in which the gravity emerges from  the
   underlying (bosonic) quantum field theory, one usually considers
   the quantity which is called the effective action $W_{eff}$. This effective
   action then describes the gravity theory that one is about to
   analyse. The crucial object in constructing the effective action is
   the trace of the heat kernel of the field operator, which in the
   case of the bosonic scalar quantum field of mass $\mu$, coupled to
   the classical gravitational background described by the metric
   tensor $g_{\mu \nu},$ is given by the d'Alembertian operator
   extended with the mass term, $\Box_{g} - \mu^2.$ On the spacetime
   with the regular geometry $g_{\mu \nu},$ whose metric in the near
   horizon region complies with the general form  (\ref{int1}),
 the trace of the heat kernel of the operator  $\Box_{g} $
    is given by the well-known Schwinger-De Witt expansion \cite{dewitt,dewitt1,dewitt2,schwinger,vassilevich, nesterov,Solodukhin:2011gn},
  \begin{equation} \label{dewitt}
    \mbox{Tr}~ e^{-s {\Box_{g}}} = \frac{1}{(4\pi)^{d/2}} \sum_{n=0}^{\infty} a_n s^{n - d/2},
  \end{equation} 
where the first few coefficients $a_n$ are given by
\begin{equation} \label{dewittcoef}
    a_0 = \int_{E} d^d x \sqrt{g}, \quad a_1 = \int_{E} d^d x \sqrt{g} \frac{1}{6} R,
\end{equation} 
\begin{equation} \label{dewittcoef1}
    a_2 =  \int_{E} d^d x \sqrt{g} \bigg( \frac{1}{180} R_{\mu \nu \rho \sigma} R^{\mu \nu \rho \sigma}
     - \frac{1}{180} R_{\mu \nu } R^{\mu \nu } +  \frac{1}{72} R^2 
     + \frac{1}{30} \Box_{g} R \bigg).
\end{equation} 
Hence, as for the first few terms in the small $s$ expansion, there come out in a sequence
 the vacuum energy term, the Einstein-Hilbert term and the higher curvature terms, respectively.

If on the other hand the conical singularity is introduced, then the resulting spacetime $E_{\alpha} $
 requires a modified small $s$ expansion for the trace of the heat kernel of the field operator
$\Box_{g} $. This is due to the fact that the Riemann curvature tensor for the space with conical singularity acquires an additional singular, delta-function like contribution \cite{starobinsky,btz1,solodukhinfursaev, Solodukhin:2011gn}, when restricted to the surface $\Sigma$.  At the same time, outside the surface $\Sigma$ it is completely identical to the curvature tensor of the regular smooth manifold $E$. Correspondingly, the small $s$ expansion (\ref{dewitt}) for the trace of the heat kernel $K(s) = e^{-s {\Box_{g}}}$ on a space with a conical singularity appropriately modifies
\begin{equation} \label{dewitt-conical}
    \mbox{Tr}_{E_\alpha}~ e^{-s {\Box_{g}}} = \frac{1}{(4\pi)^{d/2}} \sum_{n=0}^{\infty} (a_n^{reg} + a_n^{sing}) s^{n - d/2},
\end{equation} 
where the coefficients in the expansion acquire  additional
singular, i.e. surface integral contributions $a_n^{sing}$. The
components  $a_n^{reg}$  of the expansion coefficients constitute
their regular part. The first three  $a_n^{reg}$ are the same as
$a_0$, $a_1$ and $a_2$ appearing in (\ref{dewittcoef}) and
(\ref{dewittcoef1}) above, except only for the additional factor of
$\alpha$ multiplying the integrals in the expressions for
$a_0$, $a_1$, $a_2$. This factor of $\alpha$  is due to the fact that
the calculation of $a_n^{reg}$ assumes performing the integration over $E_\alpha$, instead  over $E,$
 which in turn  amounts to carrying the
integration over $E$ followed by an additional multiplication with $\alpha$.
In other words, $a_n^{reg}$ are the coefficients that would
have ruled the
expansion (\ref{dewitt-conical}) as the  sole coefficients,  if the
conical singularity had not been present at all
(that is for $\alpha =1$).

    When however the conical singularity is switched on ( $\alpha \neq 1$), the regular
   components $a_n^{reg}$ get accompanied
   by the singular components $a_n^{sing}$, which for the first
   three look as \cite{mckean,jeff,dowker,donnelly,donnelly1,fursaev, Solodukhin:2011gn}
\begin{equation} \label{dewittsingcoef}
    a_0^{sing} = 0, \quad a_1^{sing} =  \frac{\pi}{3} \frac{(1- \alpha) (1+ \alpha)}{\alpha} A(\Sigma),
\end{equation} 
\begin{equation} \label{dewittsingcoef1}
    a_2^{sing} = \frac{\pi}{18} \frac{(1- \alpha) (1+ \alpha)}{\alpha} \int_{\Sigma}  R
     - \frac{\pi}{180} \frac{(1- \alpha) (1+ \alpha) (1+ \alpha^{2})} {\alpha^{3}} \int_{\Sigma} 
    \bigg( \sum_{k=1}^{2} R_{\mu \nu} n^{k,\mu}n^{k,\nu}   
      - 2\sum_{k=1}^{2} \sum_{j=1}^{2} R_{\mu \nu \sigma \rho} n^{k,\mu}n^{j,\nu} n^{k,\sigma}n^{j,\rho}\bigg).
\end{equation} 
Here $\alpha = \beta_H/\beta_T = T/T_H$ and $A(\Sigma)$ is the area of the horizon surface. The quantities
$R, R_{\mu \nu}, R_{\mu \nu \sigma \rho} $ are respectively the curvature, Ricci tensor and the curvature tensor of the regular black hole spacetime. Since the surface $\Sigma$ is a co-dimension two hypersurface, it has two mutually orthonormal vectors $n^{k,\mu}, \; k=1,2,$ that are orthogonal to it. The indices $ \mu,\nu$ label the spacetime components of these vectors.

It has to be noted that there exists even more general expansion  for the trace of the
heat kernel on both regular as well as on the conical space. The
necessity for the generalization
 may arise for generally two reasons. One reason may be that the geometry that
 is being analysed is somewhat more involved,  meaning that in the vicinity of the horizon surface it may not simply be reduced to the direct product
 $\Sigma \times C$. This may
  e.g. be the case with the geometries describing the rotational
  spacetimes, like for example that of the Kerr black hole or the rotational BTZ,
  where the geometry in the near horizon region is no more described by (\ref{int1}).
 In these  cases there may appear additional terms in the heat kernel expansion.
In particular, the singular coefficients $a_n^{sing}, n\geq 2,$ might
suffer a mayor revision which may consist of including the extrinsic curvature effects \cite{Dowker:1994bj,mannsolodukhin,Solodukhin:2011gn}. This means
that  $a_n^{sing}$ might be affected by the additional surface integrals
of the quadratic invariants like $\sum_{j=1}^{2}\kappa^{j,\mu \nu}
\kappa^j_{ \mu \nu}$ and $\sum_{j=1}^{2}\kappa^{j} \kappa^{j},$ which would accompany the Riemann spacetime curvature terms that already exist in the expressions for $a_n^{sing}$.
Here 
 $ \kappa^{j} = g_{\mu \nu}\kappa^{j,\mu \nu}$ and $\kappa^{j}_{\mu \nu} = -\gamma^{\alpha}_{\mu} \gamma^{\beta}_{\nu} \nabla_{\alpha} n^{j}_{\beta} $ is the extrinsic curvature of the horizon surface $\Sigma$ with respect to the normal vectors $n^{j}, \; j=1,2,$ introduced above\footnote{The object $\; \gamma_{\mu \nu}= g_{\mu \nu} - n^{1}_{\mu} n^{1}_{\nu} - n^{2}_{\mu} n^{2}_{\nu} \;$ is the metric of the horizon surface \cite{mannsolodukhin,Solodukhin:2011gn}, induced by embedding it into a larger space with the metric $g_{\mu \nu}$ .}.  As observed in \cite{Dowker:1994bj, Solodukhin:2008dh} the presence of the extrinsic curvature terms
 of the above type is necessary for $a_2^{sing}$ to  manifest a general conformal invariance.
  Without such terms, $a_2^{sing}$ may at best  be invariant only under a  highly special class of conformal transformations.

Another reason for generalizing the small $s$ expansion (\ref{dewitt-conical}) may appear when one   considers the gravity theory that emerges from the underlying higher spin field theories. In that case the actual field operator for the quantum field of spin $\sigma$ is
\begin{equation}
  {\mathcal{O}}^{(\sigma)} = \Box_{g} + X^{(\sigma)}
\end{equation}
where $ \Box_{g}$ is the same d'Alembertian operator as before and $X^{(\sigma)}$ is generally a matrix depending on the spin of the field.
Correspondingly, the coefficients in the small $s$ expansion of the
trace of the heat kernel  $K(s) = e^{-s {\mathcal{O}}^{(\sigma)} }$ of the operator ${\mathcal{O}}^{(\sigma)}$
appropriately modify.

Here we shall only stick  with the scalar field operator with minimal coupling
 ($X^{(\sigma =0)} =0$). Moreover, for what concerns the analysis in this paper, only
 the coefficients $a_0^{reg},a_1^{reg},a_1^{sing}$ will be important for drawing the main
 conclusions of this article. Namely, since the main focus here is  on finding and identifying the UV divergent part of the effective action and the entanglement entropy 
of the classical BTZ  probed by the minimally coupled massless NC scalar field, only terms $a_0^{reg},a_1^{reg} $
 and $a_1^{sing}$ will be of interest. Higher terms govern the UV finite contributions to
 the effective action and the entropy in (2+1)-dimensions and thereby do not contain an inevitable  piece of information as long as the testing of the renormalization statement 
 and the comparison of the UV divergent structures obtained by different methods are the only things to look after. 
 
   Recall that in section II it was shown that a massless NC scalar field in the background of a classical spinless BTZ is mathematically equivalent to an ordinary (commutative)
   massive scalar field coupled to a classical spinning  BTZ. In this
   respect the calculation of the effective action and the
   entanglement entropy for the NC scalar field coupled to a classical nonrotating BTZ 
   is reducible to finding the effective action and the entanglement entropy for the rotational BTZ, though with different black hole parameters ($(J=0, M) \longrightarrow (J',M')$).
     Hence, in our particular model, the features induced by
   noncommutativity can be inferred by applying the conical singularity
   method onto the physical system describing the massive commutative
   scalar field, minimally coupled to a rotating BTZ black hole.

 On the other side, the term $a_1^{sing}$ will  be unaffected by the
 nonstatic nature  of the rotational black hole spacetime, as shown below.
 The nonstatic nature of the geometry in question is  not supposed to change the regular coefficients $a_0^{reg},a_1^{reg} $ either.
Naively, this can be expected on the  grounds of the general behaviour of the
 entanglement entropy, which  for the $d$-dimensional curved spacetime
 is given by the Laurent series in the UV cutoff parameter $\epsilon$,
 with the generic $n$-th term
 \cite{Solodukhin:2008dh, Solodukhin:2011gn} in the expansion
 scaling as $1/{\epsilon}^{d-2-2n}$, and with the most divergent term
 behaving as $1/{\epsilon}^{d-2}$. Therefrom it is evident that in the
 $(2+1)$-dim case only the first term will be UV divergent, and as
 being readily seen from the conical singularity method, this one draws its
 origin from the term in the heat kernel expansion that contains
 the singular coefficient $a_1^{sing}$. This is why to this purpose
 the higher coefficients can be ignored. It still remains to give a
 more direct argument as to
 why  this coefficient will  stay unaffected by the intrusion of the extrinsic curvature terms that
 may enter the formulas for the singular coefficients, owing to the nonstatic nature of the actual
 geometry (in the present case the rotational BTZ). 

  In order to put forth the statement on insensitivity of $a_1^{sing}$ to
the rise of the rotational character of the black hole spacetime
 with the horizon $\Sigma$,
 one may recall \cite{Dowker:1994bj}
 that for some general metric having the conical
singularity located at the entangling surface $\Sigma$, the only source of modification in
the singular terms in the small $s$ expansion (\ref{dewitt-conical})
can come from the extrinsic curvature of $\Sigma$  (see the discussion above).
 Moreover, since the coefficient $S_{d-2-2n},$ that stands next to the generic term 
 $1/{\epsilon}^{d-2-2n} $ ($n$-th in a row, see \cite{EEexp})  in the UV divergent part of the entanglement
 entropy expansion, cannot depend on the direction of vectors normal to
 $\Sigma$, the components $\kappa^{j,\mu \nu} $ of the  extrinsic curvature
  may appear in $S_{d-2-2n}$ only with even powers. This is the reason \cite{Solodukhin:2011gn},\cite{Solodukhin:2008dh}
  why the general coefficient $S_{d-2-2n}$ has the form 
$\sum_{k+j=n} \int_{\Sigma} R^{k} {\kappa}^{2j}$, where $R$ and
 $\kappa$ represent symbolically the components of the spacetime curvature
 tensor, e.g. the components of the extrinsic curvature of $\Sigma$, respectively.
Since the leading divergent term ($n=0$),  $S_{d-2}/ {\epsilon^{d-2}}$ in the
entropy originates from the term with $a_1^{sing}$ in the heat kernel expansion, 
it is clear that no change in the $a_1^{sing}$
coefficient is possible due to the extrinsic curvature.
 Therefore, we may conclude that in finding the UV divergent structure of the entanglement
 entropy for the model considered here, we can rely on the expressions given
 in (\ref{dewittcoef}) and (\ref{dewittsingcoef}).



Once having the heat kernel expansion, the effective action on the space with conical singularity is given as 
\begin{equation}
  W (\alpha) = -\frac{1}{2} \int_{\epsilon^2}^{\infty} \frac{ds}{s} {\mbox{Tr}}_{E_\alpha} e^{-s \Box_{g} }.
\end{equation}
The entanglement entropy is calculated by using the replica trick
\begin{equation} \label{replicatrick}
  S = {\bigg( (\alpha \partial_{\alpha} - 1) W(\alpha) \bigg)}_{\alpha = 1}.
\end{equation} 
If the scalar field has the mass $\mu,$ then the corresponding
effective action can be written in terms of the trace of the heat
kernel (\ref{dewitt-conical}) in the following way
\begin{equation}
  W (\alpha) = -\frac{1}{2} \int_{\epsilon^2}^{\infty} \frac{ds}{s}
 \bigg( {\mbox{Tr}}_{E_\alpha} e^{-s \Box_{g} } \bigg) ~ e^{-s \mu^2}.
\end{equation}

\section{Entanglement entropy and its divergent structure  for the NC scalar field in
  the classical BTZ background from the conical singularity method }

We return to the model described by the equation (\ref{eomradial}), describing the NC
scalar field in the background of the classical spinless BTZ black
hole. It was shown that this system is equivalent to the massive
scalar field probing the geometry described by the metric (\ref{eqbtzmetric}).
As explained in the previous section, the entanglement entropy of this
system may be calculated by the method of conical singularity which
consists of introducing the conical defect into the spacetime (\ref{euclideanmetric}).
This conical defect is located at the horizon and has the small deficit angle
$\delta =2\pi (1-\alpha)$. The parameter $\alpha = T/T_H$ is close to $1$ and
it measures the departure of the black hole temperature from its
equilibrium temperature $T_H$. With this, the Riemann curvature tensor
acquires an additional $\delta$-function like contribution at the
horizon and the trace of the heat kernel as well as the effective
action become functions of $\alpha$. The entanglement entropy  is then
found by the replica trick (\ref{replicatrick}).

  Although the method assumes that the
geometry to which it is applied has the near horizon limit (\ref{int1}), so
that it can be represented as $\Sigma \times C$,
and (\ref{euclideanmetric}) is not of that kind, we can  still pursue the method along the
lines described in the section IV,
as long as our primary goal is merely to extract out  the UV divergent part
of the effective action and entropy in the model considered. To this purpose we
only need to consider the coefficients $a_0^{reg},a_1^{reg} $
 and $a_1^{sing}$.

Thus, by applying the heat kernel method to the field operator
 $\Box_{g}$ defined on the background (\ref{euclideanmetric}), one gets
for the effective action
\begin{eqnarray} \label{entdiv1}
 W(\alpha) & = & - \frac{\alpha}{3} \frac{1}{{(4\pi)}^{3/2}}
 \frac{1}{{\epsilon}^3} \int d^3 x \sqrt{g} +
 \frac{\alpha}{{(4\pi)}^{3/2}} (-1) \frac{1}{\epsilon}  \int d^3 x
 \sqrt{g} \bigg( \frac{1}{6} R - {\mu}^2 \bigg) \nonumber  \\
  & - & \frac{\alpha}{{(4\pi)}^{3/2}} (\sqrt{\Lambda_{IR}} -
 \epsilon) \int d^3 x \sqrt{g} \bigg[  \frac{1}{180} R_{\alpha
 \beta \mu \nu}R^{\alpha
 \beta \mu \nu} -  \frac{1}{180} R_{\alpha
 \beta }R^{\alpha
 \beta } + \frac{1}{6} \Box_{g} \bigg( \frac{1}{5} R - {\mu}^2
 \bigg) + \frac{1}{2}  {\bigg( \frac{1}{6} R - {\mu}^2
 \bigg)}^2   \bigg] \nonumber  \\
  & + &  \frac{1}{{(4\pi)}^{3/2}} (-1) \frac{1}{\epsilon}
 \frac{\pi}{3} \frac{1 - {\alpha}^2}{\alpha} A(\Sigma)  \\
 & - & \frac{1}{{(4\pi)}^{3/2}} (\sqrt{\Lambda_{IR}} -
 \epsilon) \bigg[  \frac{\pi}{3} \frac{1 - {\alpha}^2}{\alpha}
 \int_{\Sigma}  \bigg( \frac{1}{6} R - {\mu}^2 \bigg)  \nonumber \\
 & - & \frac{\pi}{180}  \frac{1 - {\alpha}^4}{{\alpha}^3}  \int_{\Sigma}
  \bigg( \sum_{k=1}^{2} R_{\mu \nu} n^{k,\mu}n^{k,\nu}   
       - 2\sum_{k=1}^{2} \sum_{j=1}^{2} R_{\mu \nu \sigma \rho} n^{k,\mu}n^{j,\nu} n^{k,\sigma}n^{j,\rho}\bigg) + f(\alpha)\int_{\Sigma} 
        {\mathcal{O}} ({\kappa}^k {\kappa}_k, {\kappa}^{k, \mu \nu} {\kappa}_{k, \mu \nu} ) \bigg],  \nonumber
\end{eqnarray}
where the $\Lambda_{IR}$ is some conveniently chosen infrared cutoff and
 $n^{k,\mu}, \; k=1,2,$ are two mutually orthonormal vectors that are
 orthogonal to the horizon surface $\Sigma$ of the equivalent spinning
 BTZ. As stated earlier, in the effective action there may also appear the additional extrinsic curvature terms of the horizon surface  $\Sigma$ due to the nonstatic nature of the geometry. They are indicated in the above expression within the last term, where $f(\alpha)$ is some generic function of the parameter $\alpha$, with the property $f(\alpha =1) = 0,$ whose exact form is not essential for the forthcoming analysis.
The  gravitational  action that emerges from the underlying quantum
scalar field theory and which contains  the  divergent  and  finite  parts is  given  by  the regular
 part  of  the  heat  kernel  expansion, for which purpose we have to
 set  $\alpha = 1$ in (\ref{entdiv1}),
\begin{eqnarray} \label{entdiv2}
 W_{eff} & = & - \frac{1}{3} \frac{1}{{(4\pi)}^{3/2}}
 \frac{1}{{\epsilon}^3} \int d^3 x \sqrt{g} -
 \frac{1}{{(4\pi)}^{3/2}} \frac{1}{\epsilon}  \int d^3 x
 \sqrt{g} \bigg( \frac{1}{6} R - {\mu}^2 \bigg)        \\
  & - & \frac{1}{{(4\pi)}^{3/2}} (\sqrt{\Lambda_{IR}} -
 \epsilon) \int d^3 x \sqrt{g} \bigg[  \frac{1}{180} R_{\alpha
 \beta \mu \nu}R^{\alpha
 \beta \mu \nu} -  \frac{1}{180} R_{\alpha
 \beta }R^{\alpha
 \beta } + \frac{1}{30} \Box_{g}  R  + \frac{1}{2}  {\bigg( \frac{1}{6} R - {\mu}^2
 \bigg)}^2   \bigg]. \nonumber  
\end{eqnarray}

BTZ solution, despite sharing many similarities with ordinary
$(3+1)$-dimensional black holes, it has important differences as well,
rooted in the simplicity of $(2+1)$-dimensional gravity. Therefore,
due to the fact that in three spacetime dimensions, the full curvature
tensor is completely determined by the Ricci tensor,
\begin{equation} \label{rmunu}
  R_{\mu \nu \rho \sigma } = g_{\mu \rho} R_{\nu \sigma} + g_{\nu \sigma} R_{\mu \rho}
               - g_{\nu \rho} R_{\mu \sigma} - g_{\mu \sigma} R_{\nu \rho}
             - \frac{1}{2} ( g_{\mu \rho} g_{\nu \sigma} - g_{\mu \sigma} g_{\nu \rho})R, \nonumber
\end{equation}
the corresponding terms in (\ref{entdiv1}) and (\ref{entdiv2}) should be appropriately modified in accordance with that.
Nevertheless, as already announced earlier, we shall only be interested
in the divergent part of the effective gravitational action,
\begin{equation} \label{entdiv3}
  W_{div} (\epsilon) \equiv - \frac{1}{3} \frac{1}{{(4\pi)}^{3/2}}
 \frac{1}{{\epsilon}^3} \int d^3 x \sqrt{g} -
 \frac{1}{{(4\pi)}^{3/2}} \frac{1}{\epsilon}  \int d^3 x
 \sqrt{g} \bigg( \frac{1}{6} R - {\mu}^2 \bigg),
\end{equation}
as one of our main goals is to test the renormalization hypothesis within the particular NC scalar field model presented in section II. This task includes the renormalization of the bare gravitational action $W_{gr} (G_B, \Lambda_B),$ with $G_B$ and $\Lambda_B$ being the bare couplings, that is the bare Newton's gravitational constant and the bare cosmological constant, respectively. The standard way of treating the UV divergences in the action is to absorb them into a redefinition of the couplings.
While the UV divergent part in the gravitational action has already been isolated, it still remains to set up the renormalization condition that will sweep away the divergences. To this purpose, one proceeds as follows.
Taking into account that the bare and the renormalized gravitational actions are given by
\begin{eqnarray} \label{entdiv4}
 W_{gr} (G_B, \Lambda_B) & = &  \int d^3 x \sqrt{g} \bigg[-
 \frac{1}{16\pi G_B} (R+ 2\Lambda_B)
 \bigg], \nonumber  \\
 W_{gr} (G_{ren}, \Lambda_{ren}) & = &  \int d^3 x \sqrt{g} \bigg[ -
 \frac{1}{16\pi G_{ren}} (R+ 2\Lambda_{ren})    \bigg] \nonumber  
\end{eqnarray}
and that the  UV  divergences  in  the  effective  action computed  by  the  heat  kernel  method read as
\begin{equation} \label{entdiv5}
 W_{div} (\epsilon) = - \frac{1}{3} \frac{1}{{(4\pi)}^{3/2}}
 \frac{1}{{\epsilon}^3} \int d^3 x \sqrt{g} -
 \frac{1}{{(4\pi)}^{3/2}} \frac{1}{\epsilon}  \int d^3 x
 \sqrt{g}   \frac{1}{6} R + \frac{{\mu}^2}{{(4\pi)}^{3/2}} \frac{1}{\epsilon}  \int d^3 x \sqrt{g},
  \nonumber
\end{equation}
the renormalization of the effective action can be carried out by imposing the requirement
\begin{equation} \label{entdiv6}
  W_{gr} (G_B, \Lambda_B) + W_{div} (\epsilon) = W_{gr} (G_{ren}, \Lambda_{ren}).
\end{equation}
This requirement plays the role of the renormalization condition and  it leads to the redefinition of the couplings
\begin{eqnarray} \label{entdiv7}
  \frac{\Lambda_B}{G_B} +  \frac{1}{3 \sqrt{\pi} } \frac{1}{{\epsilon}^3}
  - \frac{{\mu}^2}{\sqrt{\pi}}  \frac{1}{\epsilon} =  \frac{\Lambda_{ren}}{G_{ren}},   \\
   \frac{1}{G_B} + \frac{1}{3 \sqrt{\pi}} \frac{1}{\epsilon}  = \frac{1}{G_{ren}}. 
\end{eqnarray}
It should be noted that the noncommutativity scale $a$ enters the first condition through the mass $\mu$ of the scalar probe, which was given by the third relation in Eq. (\ref{8}). Eventual impact of the scale $a$ on the cosmological constant  would  certainly be an interesting issue to study and would perhaps deserve a special analysis, but it  is beyond the scope of the present paper.

In order to test the renormalization statement, it is necessary to know how the entropy looks like for the case of the NC scalar field model of section II. Hence, 
applying the replica trick to the effective action (\ref{entdiv1}),
\begin{equation} \label{entdiv7}
  S = (\alpha {\partial}_\alpha -1) W(\alpha) \big |_{\alpha = 1},
\end{equation}
one gets the following result for the entanglement entropy 
\begin{eqnarray} \label{entdiv8}
  S = \frac{A(\Sigma)}{12 \sqrt{\pi}} \frac{1}{\epsilon} & + & \frac{1}{12 \sqrt{\pi}}
 (\sqrt{\Lambda_{IR}} - \epsilon) \bigg[ \int_{\Sigma}
   \bigg( \frac{1}{6} R - {\mu}^2 \bigg)  \\
   & - & \frac{1}{30}  \int_{\Sigma}
 \bigg( \sum_{k=1}^{2} R_{\mu \nu} n^{k,\mu}n^{k,\nu}   
        - 2\sum_{k=1}^{2} \sum_{j=1}^{2} R_{\mu \nu \sigma \rho} n^{k,\mu}n^{j,\nu} n^{k,\sigma}n^{j,\rho}
    \bigg)  + 
       (\alpha \partial_{\alpha} f(\alpha))_{\alpha =1} \int_{\Sigma} 
               {\mathcal{O}} ({\kappa}^k {\kappa}_k, {\kappa}^{k, \mu \nu} {\kappa}_{k, \mu \nu} )
     \bigg].  \nonumber
\end{eqnarray}
This is the entanglement entropy for the massless NC scalar field minimally coupled to the classical nonrotational BTZ geometry (\ref{btzmetric}). It is calculated by applying the heat kernel method for the spaces with conical singularity onto the mathematically equivalent system that consists of the ordinary (commutative) massive scalar field probing the rotational BTZ geometry (\ref{eqbtzmetric}). As it can be seen from  Eq.(\ref{entdiv8}), this entropy is a UV divergent quantity.
The leading (and at the same time the only) UV divergent term in the entanglement entropy, obtained by
the heat kernel method on the space with conical singularity, is given by
\begin{equation} \label{entdiv9}
  S_{div} (\epsilon) =
  \frac{A(\Sigma)}{12 \sqrt{\pi}} \frac{1}{ \epsilon },
\end{equation}
showing that it scales linearly with the UV cutoff parameter $\epsilon$. The horizon area
 $A(\Sigma)$ is here determined by
\begin{equation}
A(\Sigma) = 2 \pi r_+^{'} = 2\pi \sqrt{\frac{M' l^{2}}{2} \bigg(1 + \sqrt{1- \frac{J'^{2}}{M'^{2} l^{2}}} \bigg)}, 
\end{equation}
with the mass $M'$ and the angular momentum $J'$ calculated in (\ref{mprimejprime}).
It is worthy to note that the same result for the UV divergent contribution to the entanglement entropy is obtained in \cite{Mann:1996ze} where the heat kernel on the conical BTZ geometry is constructed by  solving exactly the heat equation on a maximally symmetric constant curvature space and then utilising the Sommerfeld formula \cite{sommerfeld} to obtain the heat kernel with the required periodicity. The heat kernel method for $\mbox{AdS}_3$ spaces was also considered in \cite{David:2009xg}.

The next to leading  term in the Eq.(\ref{entdiv8}) is already UV finite. 
We point out that this same UV divergent structure is exhibited by the entropy that was calculated  for the same NC model, but within the framework of the 't Hooft's brick wall method of section III.  The leading term in the entropy obtained there was also linearly divergent in $\epsilon,$ while the next following term was UV finite (see Eq.(\ref{entropymain})).

 Beside comparing the brick wall result in Eq.(\ref{entropymain}) with that in (\ref{entdiv9}), the potentially interesting conclusions may also be drawn  by 
  confronting the result (\ref{entropymain}) with that in \cite{Ho:1998du}, where the entropy of the quantized scalar field in the background of a rotating BTZ black hole has been analysed within the brick wall model, by distinguishing 
  explicitly between the contributions to the entropy that are coming from the superradiant and nonsuperradiant modes. What has been shown in \cite{Ho:1998du}  is that, although both of these contributions to the entropy
  have  the subleading logarithmically divergent terms, these subleading terms in the superradiant and nonsuperradiant parts come with the opposite sign, implying that in the total entropy they cancel with each other.
  Moreover, the leading terms in the superradiant and nonsuperradiant contributions are exactly the same, so that in the total entropy they double. Consequently, the UV divergent structure of the total entropy for the quantized scalar field 
   in the rotating BTZ background geometry, according to \cite{Ho:1998du}, is determined only by a linearly  UV divergent term which, in addition, is proportional to the area of the event horizon. This result, beside being consistent with that in \cite{Mann:1996ze}, where the entanglement entropy has been
     analysed for the same physical situation, though within the framework of the heat kernel method on spaces with conical singularity, it is also consistent with the result (\ref{entropymain}) and the  main tenet of the present paper. 
     
     In order to explain the potentially interesting implications of the above described similarities
     between  the results obtained here and in the literature, it is useful to note that the result (\ref{entropymain}) has been calculated by applying the brick wall method 
     to the case of the nonrotational BTZ geometry, though in the presence of noncommutativity.
     Contrary to that, the entropy (\ref{entdiv9}), as well as the corresponding results in \cite{Mann:1996ze} and \cite{Ho:1998du}, have been obtained by analysing a basically rotational BTZ geometry, thus clearly indicating  that the results (\ref{entropymain}) and (\ref{entdiv9})
     refer to two 
     different physical situations. 
     Nevertheless, although pertaining to two physically different settings, these results
     anyway appear to be structurally  equivalent.
     By equivalent it is meant that the entropy of each one of them is  proportional to the area of the event horizon and in addition they have the same UV divergent structure, which is characterized by the linear divergence in $\epsilon$. This observation may be found useful when studying some specific rotational geometry and trying to simplify the analysis by reducing
       the problem to a more simple, but equivalent one.
            It may  be found useful particularly in the light of the observation
   that a rotational black hole background geometry appears to be fundamentally different from the nonrotational one,   at least as far as the quantization of the matter fields is
      concerned \cite{Frolov:1989jh},\cite{Matacz:1993hs}.
            Even more so with regard to the entropy calculation since,
       along the standard  nonsuperradiant modes, in these new circumstances one also has to take care
       about the superradiant modes.
The presence of these superradiant modes makes
        the case with a rotational geometry fundamentally different and accordingly more involved when compared to the case with a nonrotational geometry.
            It is also likely that these superradiant
      modes that call for a special care in the brick wall model in fact mirror the extrinsic
      curvature effects that one encounters in studying the
 rotational geometries within the framework of the conical singularity method.
 
 Nonetheless, based on the equivalence between the nonrotational geometry in
       the presence of noncommutativity and the rotational geometry, the feature  that was explicitly exhibited through the mapping (\ref{8}),
        it turns up that a noncommutative spacetime  may provide a suitable medium
        in which  the bridging between these two 
         physically different settings might be possible to enforce, by
                 connecting and correlating their respective parameters.
                In particular, this  might help  to reduce
      one situation that is technically more  involving (a rotational one) to  another
      which is less involving (a nonrotational one), a very feature which might appear as beneficial from the practical and calculational point of view.
      
      From the present results and from the results known in the literature, it is clear that the BTZ geometry, no matter if it  is rotational or nonrotational,
     gives rise to an entropy that  is in the leading order characterized  by a linearly UV divergent contribution of the Bekenstein-Hawking type. Thereby, in the leading order the entropy is proportional to the area
     $A= 2\pi r_+$ of the event horizon, where $r_+= l\sqrt{M}$ for the nonrotational case, while for the rotational case the radius of the outer horizon gets modified by the interference of the angular momentum.  
      Moreover, a brief look at  Eq. (\ref{entropymain}) shows that the presence of noncommutative setting does not bring any change to this conclusion.
      As a matter of fact, a closer inspection of Eq. (\ref{entropymain}), specifying the entropy of the nonrotational BTZ in the presence of noncommutativity,
      leads to an interpretation of noncommutativity as giving rise to a stretch/shrinkage of the event horizon.
             Likewise, an interesting observation  may be drawn by making an explicit comparison of this result with the result (\ref{entdiv9}) or  with the formula (31)
        in \cite{Ho:1998du} or the analogous one in \cite{Mann:1996ze} (the latter three  
       being basically the same up to distinct numerical prefactors).
        Based on this comparison a common picture emerges in which 
    the noncommutative contribution in (\ref{entropymain}) appears as if it  has been soaked up into an effective angular momentum which in turn gives rise to a stretch of the event horizon.

 

We now turn the attention toward the problem of  validation of the renormalization statement within the framework of the NC scalar field model discussed in this article. In this respect, note
that the  divergency in the entanglement entropy, identified in
(\ref{entdiv9}), can be  removed  by  the standard  renormalization
procedure, during which the divergency becomes reabsorbed within the redefined
coupling constants. To this end, the bare and the renormalized entropy can
respectively be written as
\begin{equation} \label{entdiv10}
  S(G_B) = \frac{A(\Sigma)}{4G_B}, \qquad  S(G_{ren}) =
  \frac{A(\Sigma)}{4G_{ren}}.
\end{equation}
The renormalization condition,
\begin{equation} \label{entdiv11}
  S(G_B) + S_{div} (\epsilon) = S(G_{ren}), 
\end{equation}
together with the relation (\ref{entdiv9}), then leads to the renormalization of the Newton's gravitational constant,
\begin{equation} \label{entdiv12}
   \frac{1}{G_B} + \frac{1}{3 \sqrt{\pi}} \frac{1}{\epsilon}  = \frac{1}{G_{ren}},
\end{equation}
in the manner as first  proposed in \cite{Susskind:1994sm}.
Herefrom, it is readily seen that the  same  renormalization  condition  which  removes  divergences  in  the  effective  action,
also  renormalizes  the  entanglement  entropy. Hence, the renormalization statement  for the 
 particular model considered in this article, describing  NC scalar
 field coupled to the  classical BTZ geometry has been validated explicitly.

Upon utilising the renormalization condition to remove the divergences in the entanglement entropy, a relation $\frac{1}{G_{ind}} \sim \frac{1}{3 \sqrt{\pi} \epsilon}$ naturally arises,
 making for a precise balance between the induced
gravitational constant $G_{ind}$ and the entanglement entropy, so that
the entanglement entropy appears to be precisely equal to the Bekenstein-Hawking
entropy, expressed in terms of the induced gravitational constant.
As already observed, this result is congruent with  \cite{Mann:1996ze} and \cite{Ho:1998du}, when restrained to the leading order.

  Interestingly enough, a study of a $2$-dimensional conformal field theory (CFT) within the framework of the holography and AdS/CFT correspondence may lead to the same conclusion \cite{Calabrese:2004eu,Hubeny:2007xt}. That is, the information on the holographic entanglement entropy  in the bulk of $\mbox{AdS}_3$
   spacetime may be obtained by studying the dual CFT on the $2$-dimensional boundary which has a topology of a cylinder. The corresponding entanglement entropy \cite{Cadoni:2009tk,Cadoni:2007vf} for the thermal $2$-dim CFT on the cylinder for a spacelike slice of length $ \; 2\pi l \; $ is
  \begin{equation} \label{entcft}
    S_{CFT}   = \frac{l}{4G_{ind}} \ln \bigg[  \frac{l^2}{\pi^2 (r'_+ + r'_-)(r'_+ - r'_-)}
       \sinh \frac{\pi (r'_+ + r'_-)}{l} \sinh \frac{\pi (r'_+ - r'_-)}{l} \bigg]. 
     \nonumber
  \end{equation}
  In the above expression the renormalization was already undertaken by subtracting the vacuum contribution coming from the left and right movers describing the rotational BTZ.
  The macroscopic, that is the large temperature limit, $\; r'_+ >> l \;$ and $ \; r'_+ >>  r'_- \; $
  then   gives 
   $ \; \frac{\pi r'_+}{2G_{ind} } \; $ for the leading contribution, that is the Bekenstein entropy
    (\ref{entdiv9}).

\section{Final remarks}

 In the present paper we have considered the NC scalar field model coupled to the classical nonrotational BTZ geometry. For this particular model we have calculated the entropy within the two different frameworks, one being that of the 't Hooft's brick wall model and the other one being that of the heat kernel method developed for the spaces with conical singularity and then we compared the results of these two approaches. When using the heat kernel method  in particular, we have relied on the small $s$ expansion for the trace of the heat kernel of the actual field operator $ \Box_{g},$ rather then on the exact solution of the corresponding heat equation. A comparison of the results for the entropy obtained from these two different approaches shows that they both predict the identical UV divergent structure for the entropy, with the leading term being linearly divergent in the UV cutoff parameter $\epsilon,$ and the next to leading order term being UV finite, as well as the rest of the expansion for the entropy. The second goal of the paper was to test the renormalization statement for the NC model considered. Here we found  that the  same  renormalization  condition  that  removes  divergences  in  the  effective  action is also responsible for the removal of the divergencies
 in  the  entanglement  entropy. Hence, the renormalization statement  for the 
  particular model  describing  the NC scalar field in the background of  the  classical BTZ geometry has been validated.
 
 In carrying out the analysis we have utilised the exact mathematical equivalence between the noncommutative model considered and the equivalent model consisting of the massive commutative scalar field coupled to a spinning BTZ geometry.
This mathematical equivalence  itself has an interesting and
novel physical interpretation. Namely, it gives rise to 
 a novel view on noncommutativity, which emerges from our analysis, by assigning it a role of a mass generating agent, as well as a driving force that lies behind the black hole spin.

Furthermore, there is an interesting observation that can be made regarding the effect of noncommutativity on the divergent character of the entropy.
To wit, we have seen that the mere effect of noncommutative nature of spacetime  was to shrink or stretch the event horizon of a black hole, seemingly without any impact whatsoever
on the UV structure of the entropy, neither changing it for the better nor worsening it further. However, there may be an indirect impact, as may be seen from the following line of reasoning.

As it is well known, when the UV cutoff parameter $\epsilon$ approaches $0,$ the entropy blows up.
However, the presence of noncommutativity implicitly presumes the existence of the minimal distance scale $a$ beyond which it is not possible to go.   
 This in turn means that the limit $\epsilon \rightarrow 0$ cannot be applied to the full extent, since otherwise  $\epsilon$ would at some stage cross the barrier set out by the noncommutative scale $a$.
By way of,  $\epsilon$ at best can reach $a,$ but cannot go beyond, i.e. it cannot reduce further.
Due to the existence of this  natural length  barrier, one may argue that the noncommutative nature of spacetime provides a setting which avoids the problem with the divergences in the entropy
because, no matter how small $a$ may be, the entropy, though very large, will still remain finite.

As a matter of fact, the UV cutoff $\epsilon$ can be fixed by relating it with the NC scale parameter $a$ as\footnote{So far we were carrying the study in the units where 8G = 1.
 For the purpose of the remaining analysis we switch to the
standard unit system, which in turn formally corresponds to putting 8GM everywhere in place of M.} 
  \begin{equation}
 \epsilon = \frac{3\zeta(3)}{2 \pi^{3}} G \bigg[ 1+ a \bigg( \frac{8\pi}{3}\beta \frac{\sqrt{8GM}}{l} \frac{\zeta(2)}{\zeta(3)}  - \frac{1}{2} \bigg) \bigg],
 \end{equation}
 with the latter bound stemming from the comparison of (\ref{entropymain}) with (\ref{entdiv9}) and by utilising (\ref{entdiv12}). Therefrom it is readily seen that $a$ pushes the brick wall cutoff $\; h = \frac{\sqrt{8GM}}{2l} \epsilon^{2} \; $ slightly below or above the classical ' t Hooft's bound, depending on the value of the black hole mass $M$ and the sign of the parameter $\beta$. 


Finally, it should be noted that in the first place, when the brick wall method was used, it was applied to a static black hole, though in the presence of noncommutativity. On the contrary, when the conical space approach was used later on, the full equivalence with the model of rotational BTZ geometry was utilised and the method was applied directly to this rotational case.
 Besides that both results are being consistent with each other, they also turn out to be consistent
  with the results established in the earlier literature. Here we
 specifically  have in mind the leading contribution to the entropy that  was
 obtained for the rotational BTZ geometry within the brick wall method
 \cite{Ho:1998du}, and also the leading contribution to the entropy that was obtained
   for the same geometry, yet analysed within the framework of the
 heat kernel method on
 spaces with conical singularity \cite{Mann:1996ze}.
 Moreover, a direct comparison of the two results (\ref{entropymain}) and (\ref{entdiv9})
attributes to noncommutativity a role of a medium that mediates in the
 process of stretching of a horizon  through the
 appearance of an  effective black hole angular momentum,
 which seemingly  takes over the whole noncommutativity onto itself.
 
   As the model of  NC scalar field in the background of the classical spinless BTZ
  was shown to be equivalent to the rotational BTZ geometry probed by
  a massive scalar field, our confirmation of the renormalization
  statement for the NC model may then be seen as a mere consequence of
  the similar statement as applied to the rotational BTZ. In this respect,
the conclusions presented here are in fact an indirect consequence of the
  results that are so far acquired with regard to the entropy of the rotational BTZ \cite{Ho:1998du, Mann:1996ze}.
To put it differently, they have just been rediscovered in the new context, that  of the
  noncommutative setup mingled with the ($2+1$)-dimensional gravity.
 
\noindent{\bf Acknowledgment}\\
A.S. is grateful to the members of the theory group (Mathematics and
Physics department) at the University
of Cagliari on very valuable discussions and particularly to S.Mignemi and
M.Cadoni who also provided many
thoughtfull and inspiring observations. 
 We are also grateful to  Kumar S. Gupta for giving many valuable and important comments.
Helpful comments by I.Smoli\'c are also appreciated.
This work was supported by the European Commission and the Croatian Ministry of Science, Education and Sports through grant project financed under the Marie Curie FP7-PEOPLE-2011-COFUND, project NEWFELPRO.
The work by T. J. has been fully supported by Croatian
Science Foundation under the
project (IP-2014-09-9582).



\begin{thebibliography}{99}
\bibitem{EEexp}
   $S_{n}$ is the generic coefficient that stands in front of the power $1/\epsilon^n $ in the 
   expansion of the entanglement entropy, which has the following general form 
  $$
       S = S_{d-2} / \epsilon^{d-2} + S_{d-4} / \epsilon^{d-4} +...+S_{d-2-2n} /
  \epsilon^{d-2-2n} + s_0 \ln \epsilon + \mbox{finite}.
  $$
  This expansion is the series  in terms of the UV cutoff $\epsilon$.
  The coefficients $S_{n}$ are close to the coefficients $a_{n}$ that appear in the expansion $(\ref{dewitt-conical})$, but differ from them
  by some multiplicative factor which occurs due to an additional integration   over the small adiabatic parameter $s$ that is one supposed to carry out in $(49)$.
\end{thebibliography}
\end{document}